\documentclass[preprint,aps,ams]{revtex4}

\usepackage{graphics}

\begin{document} \title{Reply to Comment on "Quantum
Measurement and Decoherence"} \author{G. W. Ford} \affiliation{Department
of Physics, University of Michigan, Ann Arbor, Michigan 48109-1120}
\author{R. F. O'Connell$^{\dagger}$}
\affiliation{Department of Physics and Astronomy, Louisiana State
University, Baton Rouge, LA 70803-4001} \date{\today } \begin{abstract}
While agreeing with our exact expression for the time dependence of the
motion of a free particle in an initial superposition state,
corresponding to two identical Gaussians separated by a distance $d$, at
temperature $T$, Gobert et al., in the preceding Comment [Phys. Rev. A
xxx], dispute our conclusions on decoherence time scales. However, the
parameters they used to generate their figures are outside the regime of
validity of our interpretation of the results and, moreover, are not of
physical interest in that they correspond to $T\approx 0$. The point is
that in their figures they have chosen the thermal de Broglie wavelength
$\lambda _{th}=\hbar \sqrt{mkT}$ to be equal to slit spacing \emph{d},
whereas we have clearly stated [in the paragraph preceding Eq. (21) of
our paper] that decoherence occurs and that our expression for the
decoherence time applies only in the limit where \emph{d} is large
compared not only with the slit width $\sigma $ but also with the thermal
de Broglie wavelength, $d\gg \lambda _{th},\sigma $. \\
\\
\\
\\
\\
\\
\noindent PACS number(s): 03.65.Yz, 05.30.-d, 05.40.Jc
\end{abstract}  \maketitle For a free
particle in an initial superposition state, corresponding to a pair of
Gaussians each of width $\sigma $ and separated by a distance $d$, at
temperature $T$, we presented an exact expression for the coordinate
probability, $P(x,t)$ \cite{ford1,ford2}. Gobert et al. (hereafter
referred to as GDA), in the preceding Comment \cite{gobert}, agree with
this result, which is valid for arbitrary temperature $T$ and arbitrary
strength of the coupling to the heat bath as measured by the dissipative
decay rate $\gamma $ . GDA also agree that their results for the
corresponding reduced density matrix are consistent with our results for
$W(q,p,t)$, the Wigner function for the probability in Wigner phase space
\cite{murakami} (we note that the Wigner distribution function is the
Fourier transform of the density matrix and contains the same information
\cite{hillery}). Based on our result for $P(x,t)$ we concluded that
decoherence without dissipation can occur for all cases of interest. GDA
dispute this conclusion. Our purpose here is to show their analysis is
flawed due to a choice of parameters that are not of physical interest in
that they correspond to $T\approx 0$. The point is that their figures
correspond to a slit spacing \emph{d} equal to the thermal de Broglie
wavelength, \begin{equation} \lambda _{\mathrm{th}}=\frac{\hbar
}{\sqrt{mkT}}. \label{cdd0} \end{equation} But we have clearly stated
[in the paragraph preceding Eq. (21) of our paper] that decoherence
occurs and that our expression for the decoherence time applies only in
the limit where \emph{d} is large compared not only with the slit width
$\sigma $ but also the thermal de Broglie wavelength, $ d\gg \lambda
_{th},\sigma $. Indeed, decoherence is fully developed even for the very
modest ratio $d/\lambda _{\mathrm{th}}$. To illustrate this we show in
Fig. 1 a plot of $P(x,t)$ [multiplied by $\sigma $ to make it
dimensionless] versus $x$ (divided by $\sigma $) at a time $t=2m\sigma
d/5\hbar \equiv t_{\mathrm{mix}}/5$ for three different values of the
ratio $ d/\lambda _{\mathrm{th}}$. The solid curve corresponds to
$d/\lambda _{ \mathrm{th}}=5$ and there we see that there is no hint of
an interference pattern, that is, decoherence has occured. The two dashed
curves, which are nearly indistinguishable, correspond to the GDA choice
$d/\lambda _{\mathrm{ th}}=1$ and the zero-temperature case $d/\lambda
_{\mathrm{th}}=0$, the zero temperature case having the slightly larger
interference amplitude. In all three cases we have neglected the coupling
to the bath (that is, set $\gamma =0$) and made the GDA choice
$d=20\sigma $. We assert that this Figure refutes those of GDA, showing
that under the conditions we have stated decoherence without dissipation
does occur. GDA ask how can it be that the interference pattern has not
disappeared in their Fig.1. The answer of course is that they have chosen
values of the parameters corresponding to an effectively zero
temperature, for which in the absence of dissipation there is of course
no decoherence. As our Fig. 1 shows, if they were to choose only a
modestly higher temperature, corresponding to $d/\lambda
_{\mathrm{th}}=5$, they would see no interference. We forbear to present
a repeat of GDA's Fig. 1, since for zero temperature the figure would be
indistinguishable from theirs, while for $ d/\lambda _{\mathrm{th}}=5$
the figure would be utterly featureless, consisting of two Gaussians
propagating independently without interference. Perhaps a numerical
illustration might be helpful. Since the magnitude of the de Broglie
wavelength $\lambda_{th}$ plays a key role in the analysis, we use
(\ref{cdd0}) to write \begin{equation} \lambda^{2}_{\mathrm{th}}=(5\cdot
2\times 10^{-21}cm)^{2}\left( \frac{1gm}{m} \right) \left(
\frac{300K}{T}\right). \label{cdd1} \end{equation} As an illustrative
example, Zurek, in his oft-cited paper \cite{zurek}, chooses $m=1gm$,
$d=1cm$ and $T=300K$, corresponding to the very large ratio $d/\lambda
_{\mathrm{th}}=2\times 10^{20}$. Put another way, the GDA choice of
$d/\lambda _{\mathrm{th}}=d$ would require a temperature of $8.1\times
10^{-39}K$! This is perhaps an extreme choice of parameters, but it
illustrates the fact that the condition $d\gg \lambda _{th}$ is not
unique to our discussion but has generally been understood to apply in
earlier discussions of decoherence. In general, the decoherence times
$\tau _{ \mathrm{d }}$ are typically much smaller than the decay time
$\gamma ^{-1}$. But this is so only under our condition $\lambda
_{\mathrm{th}}\ll d$, no matter how one defines decoherence. Next, we
turn to the question of how decoherence should be defined. As we have
stressed more than once, "-- a quantitative measure of decoherence
depends not only on the specific system being studied but also on whether
one is considering coordinate, momentum, or phase space" \cite{murakami}.
On the other hand, based on their density matrix analysis (which is
nothing more than that given previously by us for Wigner phase space
\cite{murakami} ), GDA claim that this implies that there is no
decoherence in physical space, which as we have seen is not true.
Moreover, while again recognizing that there are various ways of defining
decoherence, our definition of an attenuation coefficient $a(t)$ was such
as to ensure that $a(t)=1$ for all times in the case where $T=0$ and
$\gamma =0$, corresponding to a pure Schr \"{o}dinger cat state at all
times (as discussed in detail in Sec. VI B. of \cite{ford3}). This choice
leads to a time scale which is convenient for this system but other
definitions of decoherence time (which will also involve the temperature)
might also be considered, depending on what particular features of the
time development are of interest. For example, it is instructive to write
our general result for $P(x,t)$ in the form \cite {ford1,ford4}
\begin{eqnarray} P(x,t) &=&
\frac{1}{2\left(1+e^{-d^{2}/8\sigma^{2}}\right)}
\left\{P_{0}\left(x-\frac {d}{2},t\right)+P_{0}\left(x+\frac{d}{2}
,t\right)\right. \nonumber \\ &&{}+ \left.
2e^{-d^{2}/8w^{2}(t)}a(t)P_{0}(x,t)\cos\frac{[x(0),x(t)]xd}{
4i\sigma^{2} w^{2}(t)}\right\}, \label{cdd2} \end{eqnarray} where $P_{0}$
is the probability distribution for a single wavepacket, given by
\begin{equation} P_{0}(x,t)=\frac{1}{\sqrt{2\pi
w^{2}(t)}}\exp\left\{-\frac{x^{2}}{2w^{2}(t)} \right\}. \label{cdd3}
\end{equation} Here and in (\ref{cdd2}) $w^{2}(t)$ is the variance of a
single wavepacket, which in general is given by \begin{equation}
w^{2}(t)=\sigma^{2}-\frac{[x(0),x(t)]^{2}}{4\sigma^{2}}+s(t),
\label{cdd4} \end{equation} where $\sigma^{2}$ is the initial variance,
$[x(0),x(t)]$ is the commutator, and \begin{equation} s(t)=\left\langle
\left\{x(t)-x(0)\right\}^{2}\right\rangle , \label{cdd5} \end{equation}
is the mean square displacement. Also, $a(t)$, which can be defined as
the ratio of the factor multiplying the cosine in the interference term
to twice the geometric mean of the first two terms \cite{ford2} is given
by the following exact general formula \begin{equation}
a(t)=\exp\left\{-\frac{s(t)d^{2}}{8\sigma^{2}w^{2}(t)}\right\}.
\label{cdd6} \end{equation} As we emphasized in \cite{ford2}, it is only
for the case of Ohmic dissipation, high temperature $(kT>>\hbar\gamma )$
and $d>>\lambda_{th}$, $ \sigma$, that (\ref{cdd6}) reduces to
\begin{equation}
a(t)\rightarrow\exp\left\{-\left(\frac{t}{\tau_{d}}\right)^{2}\right\},
\label{cdd7} \end{equation} where \begin{equation} \tau
_{d}=\frac{\sqrt{8}~\sigma ^{2}}{\sqrt{kT/m}~d}, \label{cdd8}
\end{equation} the result quoted by GDA in their equation (4). But, this
result clearly does not apply for $d=\lambda _{th}$, the GDA choice of
parameters.

GDA continually refer to FLO's "-- imperfect preparation of
the initial state --". What perhaps they mean is that the initial state
of FLO is not a pure state, which is certainly the case. But a pure state
necessarily corresponds to a particle at zero temperature. FLO argue that
the picture of a particle at zero temperature suddenly coupled to a heat
bath at a high temperature $T$ is not a realistic picture of the physical
situation. The essence of FLO's paper is to show that one can, by
measurement, prepare an initial state that is entangled with the bath,
with a temperature equal to that of the bath, but with a probability
distribution identical with that of the pure state. They argue that this
is a more realistic initial state. We conclude that the comment of GDA is
simply misleading due to their unrealistic choice of parameters and we
reiterate our previous conclusion that decoherence can occur without
dissipation.

\newpage

\begin{figure}
\caption{$P(x,t)$ the coordinate probability at time $t$ (multiplied by $\sigma$
to
make it dimensionless) as a function of $x$ (divided by $\sigma$) is shown at a
time
$t=t_{\mathrm{mix}}/5$. The solid curve, which shows no hint of interference,
corresponds
to $\lambda _{
\mathrm{T}}=d/5$. The two dashed curves, which are nearly indistinguishable,
correspond to
the GDA choice of parameters: $\lambda _{\mathrm{T}}=d$ and to the zero
temperature case
$\lambda _{\mathrm{T}}=\infty $, with the zero temperature interference pattern
having a
slightly larger amplitude.}
\end{figure}


\begin{references}

\bibitem[$^{\dagger}$]{rfoc} E-mail
Address: rfoc@rouge.phys.lsu.edu

\bibitem{ford1} G. W. Ford, J. T. Lewis
and R. F. O'Connell, Phys. Rev. A {\bf{64}}, 032101 (2001), referred to
in the text as FLO.

\bibitem{ford2} G. W. Ford and R. F. O'Connell, Phys.
Lett. A {\bf{286}}, 87 (2001).

\bibitem{gobert} D. Gobert, J. von Delft
and V. Ambegaokar, Phys. Rev. A, xxx.

\bibitem{murakami} M. Murakami, G.
W. Ford and R. F. O'Connell, Laser Phys. {\bf{13}}, 180 (2003).

\bibitem{hillery} M. Hillery, R. F. O'Connell, M. O. Scully, and E. P.
Wigner, Phys. Reports {\bf{106}}, 121 (1984).

\bibitem{zurek} W. H.
Zurek, Phys. Today {\bf{44}}(10), 36 (1991).

\bibitem{ford3} G. W. Ford
and R. F. O'Connell, Am. J. Phys. {\bf{70}}, 319 (2002).

\bibitem{ford4}
G. W. Ford and R. F. O'Connell, J. Optics B {\bf{5}}, S609 (2003).
\end{references}
\end{document}